\documentclass[table]{ifacconf}

\usepackage[english]{babel}

\usepackage{enumitem,graphicx,xcolor,tabularx}
\usepackage{amsmath,amssymb, amsfonts,dsfont,cases,mathtools,bbm,bm}  

\usepackage{natbib}        
\usepackage[utf8]{inputenc}
\makeatletter
\let\old@ssect\@ssect 
\makeatother

\usepackage[hyphens]{url}
\PassOptionsToPackage{hyphens}{url}
\usepackage[colorlinks=true,linkcolor=blue!80!black,citecolor=blue!80!black,urlcolor=blue!80!black,breaklinks]{hyperref}
\makeatletter
\def\@ssect#1#2#3#4#5#6{%
  \NR@gettitle{#6}
  \old@ssect{#1}{#2}{#3}{#4}{#5}{#6}
}
\makeatother

\definecolor{gray}{gray}{0.92}

\newcommand\score{\mathrm{score}}
\newcommand\hscore{\mathrm{score}_H}

\newcommand{\btabitemize}{\vspace{-0.5cm}\begin{enumerate}[leftmargin=22pt,itemsep=0pt,label=Q\arabic*.]}
\newcommand{\etabitemize}{\vspace{-0.25cm}\end{enumerate}}

\renewcommand{\thesubsubsection}{\thesubsection.\arabic{subsubsection}}

\graphicspath{{../Figures/}}

\begin{document}
\newcolumntype{G}[1]{>{\centering\let\newline\\\arraybackslash\hspace{0pt}}b{#1}}
\newcolumntype{Z}[1]{>{\centering\let\newline\\\arraybackslash\hspace{0pt}}m{#1}}
\newcolumntype{C}[1]{>{\centering\let\newline\\\arraybackslash\hspace{0pt}}m{#1}}
\begin{frontmatter}
\title{Social interactions for a sustainable lifestyle: The design of an experimental case study\thanksref{footnoteinfo}}

\thanks[footnoteinfo]{%
This work is supported by the project Humanizing the Sustainable Smart City (\url{hiss-digitalfutures.se}) 
within Digital Futures, the Swedish Energy Authority, and IQ Samhällsbyggnad, E2B2 programme, grant agreement n. 47859-1 (cost-and energy-efficient control systems for buildings), and by the Swedish Foundation for Strategic Research-SSF-, grant agreement n. RIT17-0046 (CLAS-cybersäkra lärande reglersystem). The KTH Live-In Lab has been initiated and made possible by a donation from the Einar Mattsson-Group, whose support is kindly acknowledged. Donations from Akademiska Hus and Schneider Electric are also kindly acknowledged.}

\author[]{A. Fontan$^*$, M. Farjadnia$^{**}$, J. Llewellyn$^{***}$, C. Katzeff$^{***}$}
\author[]{M. Molinari$^{**}$, V. Cvetkovic$^{****}$, K. H. Johansson$^*$}

\address[]{Division of Decision and Control Systems,\\
$^{**}$Division of Applied Thermodynamics and Refrigeration,\\
$^{***}$Division of Strategic Sustainability Studies,\\
$^{****}$Division of Resources, Energy, and Infrastructure,\\
KTH Royal Institute of Technology, 100 44 Stockholm, Sweden, E-mail: \{angfon,mahsafa,josephll,ckatzeff,marcomo,vdc,kallej\}@kth.se}

\begin{abstract}\;
Every day we face numerous lifestyle decisions, some dictated by habits and some more conscious, which may or may not promote sustainable living. Aided by digital technology, sustainable behaviors can diffuse within social groups and inclusive communities.
This paper outlines a longitudinal experimental study of social influence in behavioral changes toward sustainability, in the context of smart residential homes. Participants are residing in the housing on campus referred to as KTH Live-In Lab, whose behaviors are observed w.r.t. key lifestyle choices, such as food, resources, mobility, consumption, and environmental citizenship.
The focus is on the preparatory phase of the case study and the challenges and limitations encountered during its setup. In particular, this work proposes a definition of sustainability indicators for environmentally significant behaviors, and hypothesizes that, through digitalization of a household into a social network of interacting tenants, sustainable living can be promoted. Preliminary results confirm the feasibility of the proposed experimental methodology.
\end{abstract}

\begin{keyword}
Sustainable behavior, experimental study, Live-In Lab, smart homes, cyber-physical-human systems, social networks.
\end{keyword}

\end{frontmatter}

\section{Introduction}\label{sec:intro}
\begin{figure}[!tbp]\centering\setlength{\tabcolsep}{0pt}
\begin{tabular}{Z{.26\textwidth}Z{.22\textwidth}}
    \includegraphics[height=4cm]{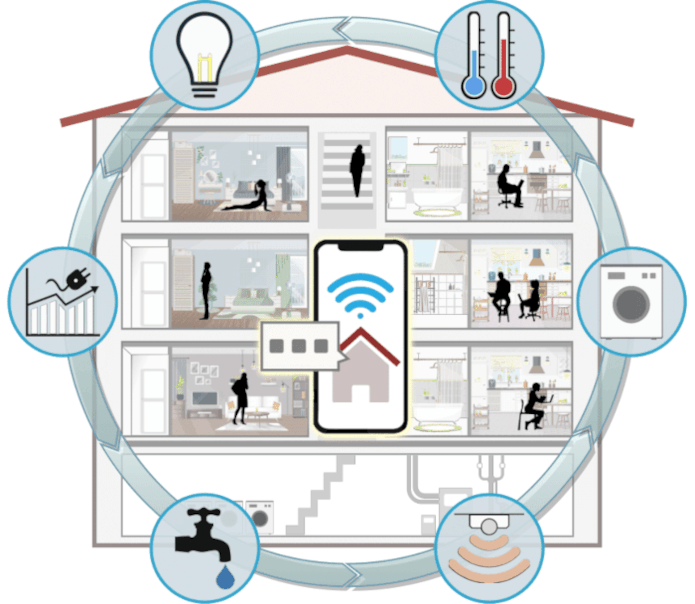}&
    \includegraphics[height=3.3cm]{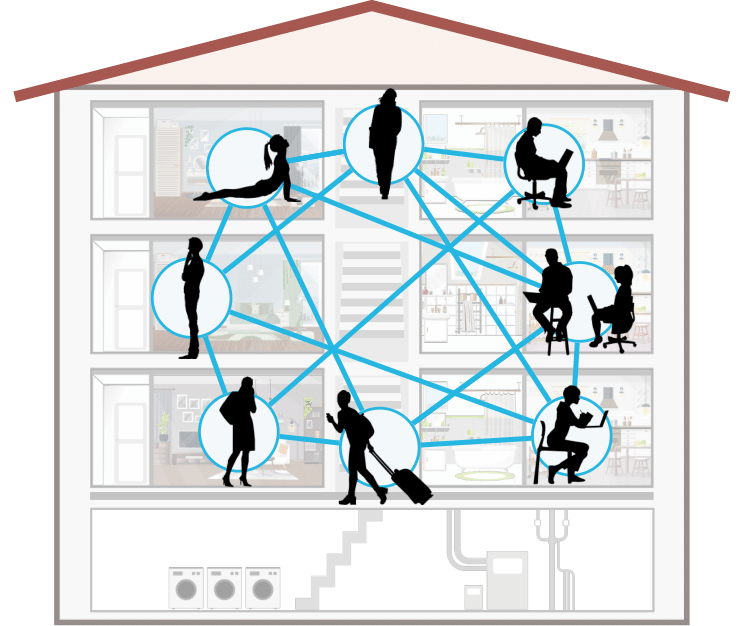} \\
    \small(a) & \small(b)
\end{tabular}
\caption{
Residential buildings are examples of complex cyber-physical-human systems, where the presence of interactions between humans and cyber-physical systems is represented not only by the reciprocal influence between tenants' behaviors and environmental conditions in residential buildings, but also by the role exerted by peer pressure. Limited comprehension of these interactions poses a challenge in model-based control designs for energy-efficiency. In this work, we use a social network to model the interactions among tenants.
(a): Smart homes as cyber-physical-human systems. (b): Smart homes as social networks.
}
\label{fig:smart-home}
\end{figure}

Recent years have seen the rapid development of the research area focused on wellbeing in smart cities, seen as complex cyber-physical-human systems (Fig.~\ref{fig:smart-home}(a)), with increasing attention on the human component and on its interplay with cyber-physical systems (\cite{CPHS2022}).
In this context, sustainable development and environmental management are of undeniable relevance for the new urban society (\cite{Bibri2017Smart,Karvonen2021Urban}) and play a significant role in motivating new societal-scale challenges for authorities (\cite{web:Sweden-sustainable,web:Swedish-energy}) and for different research communities, such as the control community (\cite{Lamnabhi-Lagarrigue2017Systems}).

\subsection{Motivation and Background}
The study we introduce in this work, called the \textit{Social KTH Live-In Lab}, focuses on smart residential homes and belongs to a broader research project whose ambition is to improve the quantitative description and predictability of human choices relevant for sustainable development related to smart cities. 
We propose an experimental study aimed at investigating if and how behavioral changes towards sustainability are affected by the collective (household), as part of a (long-term) plan to design approaches to improve building efficiency. 
The building sector is indeed critical for sustainability and the energy policy. 
It is estimated to represent approximately 30\% of the global energy consumption (\cite{Hamilton2020GlobalStatus,IEA2022}), and 17.5\% (10.9\% in residential buildings) of global energy-related emissions (\cite{Ritchie2020}); in Sweden, the residential buildings sector accounts for approximately 40\% of the energy supply (\cite{IEA2019}).

The idea to explore diverse strategies to investigate and motivate behavior change in (smart) residential homes is not novel, and approaches range from modeling household and energy use decision-making behavior (\cite{Wilson2007Residential,Peng2012Residential}), to planning ad hoc social (behavioral) interventions regarding habits (\cite{Steg2009Encouraging,Frederiks2015Household}), to designing new technologies and infrastructures, such as flexible Live-In Laboratories (\cite{Intille2006SmartPeople,Intille2006LiL,das-2020do,Spanos2020,Knutsson2016Habitation,Kalagasidis2017HSB}).
The concept we introduce in this work combines these factors and proposes a social network perspective that is, to our knowledge, novel: it frames the experimental design as a \textit{collective} (household) decision-making process with \textit{interconnected} tenants of a \textit{Live-In Lab} as the decision-makers, and its implementation relies on the collection of both qualitative (i.e., surveys distributed to participants) and quantitative (i.e., KTH Live-In Lab observations)  data.

\begin{table*}[!ht]
\caption{KPIs on Sustainability Practices.}
\label{tab:activities}\centering
\setlength\tabcolsep{2pt}\rowcolors{1}{white}{gray}
\begin{tabular}{C{.13\textwidth}|C{.7\textwidth}|C{.13\textwidth}}\hline
	\multicolumn{2}{c|}{\textbf{Sustainability indicators}}
	& \textbf{Measurability}
	\\[2pt]\hline
	\textbf{Conservation or Resources}
	& 
	\btabitemize
    \item Turn off the lights, or switch off electronic devices when leaving a room 
    \item Use the microwave or stove to warm up food instead of using the oven 
    \item Cut down on heating/decrease the temperature of the room to limit energy use 
    \item Limit the time in the shower in order to conserve energy
    \item Wait until full load to use the washing machine or dishwasher
    \item Wash clothes at temperatures $< 40^\circ C$ (compared to temperatures $\ge 40^\circ C$)
    \item Open the windows to ventilate the rooms during cold days 
	\etabitemize
	&
	Weekly Surveys, to compare with data collected at KTH Live-in Lab, see Section~\ref{sec:KTH-LiL}
	\\
	\textbf{Consumption}
	& 
	\btabitemize\setcounter{enumi}{7}
	\item Recycle the (home) waste
    \item Buy second hand items (e.g., clothes, electronic devices) instead of new ones
    \item Decide to repair an item instead of buying it new
	\etabitemize
	&
	Weekly Surveys\\
	\textbf{Food}
	& \btabitemize\setcounter{enumi}{10}
		\item Consume non-meat options (vegetarian/vegan/fish) compared to meat options 
		\item Consume non dairy options compared to dairy options
	\etabitemize
	&
	Weekly Surveys\\
	\textbf{Transportation or Mobility}
	& \btabitemize\setcounter{enumi}{12}
		\item Use public transportation instead of driving
		\item Walk, cycle, and/or use electric scooters instead of driving 
	\etabitemize
	& Weekly Surveys\\
	\textbf{Environmental citizenship}
	& \btabitemize\setcounter{enumi}{14}
    \item Watch TV programs, movies, and/or internet/social media videos about environmental issues
    \item Discuss with others outside the household about their environmental behavior\newline(Note: discuss = interact/talk, referring also to social media posts)
    \item Interact with/talk to neighbors about their environmental behavior
    \etabitemize
	&
	Weekly Surveys\\
\hline
\end{tabular}
\end{table*}
\subsection{Contributions}
The main contribution of this paper is to introduce the design of the preparatory phase of a longitudinal study spanning a 5-weeks period, whose aim is to test proposed methodology and receive feedback from a small group of participants. 
Future plan is to launch the execution phase over a 3-months period involving a large group of participants.
A longitudinal study is necessary to detect changes in behavior over time. 
Different from other studies on sustainable behavior, such as \cite{Prati2017Longitudinal}, who tested the interplay among environmental attitudes and behaviors, social identity, and pro-environmental institutional climate only twice over 2 months, in our study data collection follows a weekly schedule.

This study aims at investigating the role of social influence in environmentally significant behaviors in the context of smart homes. In particular, our approach consists in defining a \textit{sustainability score} for each tenant and observe how it dynamically changes over time.
We define social influence in terms of two sources of interaction between the occupants, which we call communication and observation, and we are interested in understanding if: (i) by discussing with neighbors, participants will change their attitudes, hence environmentally significant behaviors, hence sustainability score; and (ii) if, by observing neighbors’ behaviors (represented by the average of sustainability scores in the household) participants will change their behavior, hence sustainability score.
The hypotheses that we aim at testing can be loosely formulated as: (i) The tenants feel a strong sense of social belonging in the context of sustainable behavior, and (ii) Social influence (by communication and by observation) will predict changes in tenants' sustainability scores.

The process we propose could be explained via (theoretical) dynamical networked models for opinion dynamics and decision-making. Using digital tools, which allow virtual communication between individuals, we can map a smart building into a social network of interacting tenants (the decision-makers, Fig.~\ref{fig:smart-home}(b)) exchanging opinions with each other, and we can represent tenants' environmentally significant behaviors in terms of their attitudes (opinions) and lifestyle choices (decisions). 
Tenants' attitudes and decisions can be represented by state variables and their time evolution can be modeled as a networked dynamical system. 
A review of common models in the literature of opinion dynamics is offered in \cite{Proskurnikov2017Tutorial,Anderson2019Recent}. In particular, the design proposed in this paper is inspired by ``two-scales'' and ``two-layers'' modeling approaches (\cite{Tian2018Opinion,Wang2021Concatenated,Zino2020Twolayer}). The idea is that, even in the presence of discrepancies or disagreements of opinions in the initial state, coupling by sequential concatenation of communication (i.e., exchange of opinions) followed by observation (i.e., of neighbor's actions) may lead to the adoption of a novel social norm. In the smart homes context that is to say, even if tenants may have conflicting beliefs or behaviors on sustainability practices, sequential interactions may help achieve a common understanding and sustainable conduct at a household level.
This is a direction that we intend to explore in future works, inspired by previous examples in the literature of collective decision-making over real-world social networks such as, e.g., \cite{Fontan2021Signed}, describing government formation processes in parliamentary democracies, and \cite{Bernardo2021Paris}, describing the achievement of the 2015 Paris Agreement on climate change.

The challenges involved in the experimental design include: (i) providing a definition of environmentally significant behaviors in the context of sustainability in residential homes, and (ii) identifying an adequate measurement system, to be implemented while taking into account the availability of, and effort required from, the participants.
Various definitions of environmentally significant behavior have been proposed in the literature (\cite{Stern2000Toward,Markle2013PEBS,Steg2009Encouraging}). The definition that we adopt is based mostly on the works by P. Stern and of G. L. Markle, who define behavior in terms of its environmental impact, by determining the most significant consequences \textit{first}, \textit{and then} the activities responsible for those consequences \cite[p.~907]{Markle2013PEBS}. The main advantage of this approach is the identification of a set of activities for which a correlation ``behavior $\sim$ impact'' is established. 
In her work, Markle identifies a behavioral scale with 19 items (i.e., activities) grouped into four dimensions. Adapting her scale to our research study, we define environmentally significant behaviors in terms of sustainability practices grouped around five dimensions, namely, resources, consumption, food, mobility, and environmental citizenship (Table~\ref{tab:activities}).
The resulting 17-items questionnaire is to be distributed weekly to participants; importantly, the outcome of this survey allows for the calculation of a \textit{sustainability score} for each tenant, which we can monitor over time to observe changes in behavior. Meanwhile, the data on resources (energy and water consumption) will be compared and validated with directly observable measurements from the sensors placed at the KTH Live-In Lab. 

\subsection{Outline}
The paper is organized as follows: Section~\ref{sec:setup} introduces the experimental setup and gives a brief overview of the KTH Live-In Lab and its tenants; Section~\ref{sec:LiLSocial} presents the design of the research study; Section~\ref{sec:results} presents preliminary results; finally, Section~\ref{sec:conclusion} offers conclusive remarks.

\section{Experimental setup}\label{sec:setup}
The experimental campaign is run at the KTH Live-In Lab\footnote{\url{https://www.liveinlab.kth.se/en/infrastruktur/testbed-infrastructure}}, a platform of building testbeds that enables, among other opportunities, to carry out real-life experiments in buildings. 
It is an example of CPHS environment, which combines the KTH Live-In Lab buildings and datapool (see Section~\ref{sec:KTH-LiL}), and the KTH Live-In Lab tenants (see Section~\ref{sec:tenants}).

\subsection{KTH Live-In Lab}\label{sec:KTH-LiL}
\begin{figure*}[!t]\centering\setlength{\tabcolsep}{0pt}
\begin{tabular}[t]{C{.38\textwidth}C{.52\textwidth}}
\includegraphics[height=2.9cm,trim={0 0 0 0.5cm},clip]{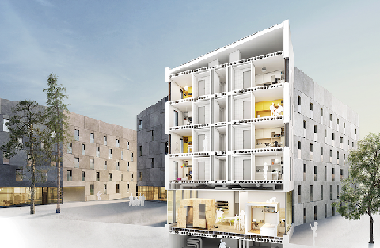}\newline\vspace{0.2cm}
\includegraphics[height=3.9cm,trim={0 0 0 0.5cm},clip]{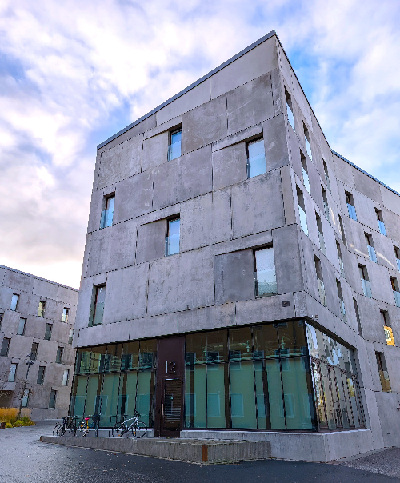}\;
\includegraphics[height=4cm]{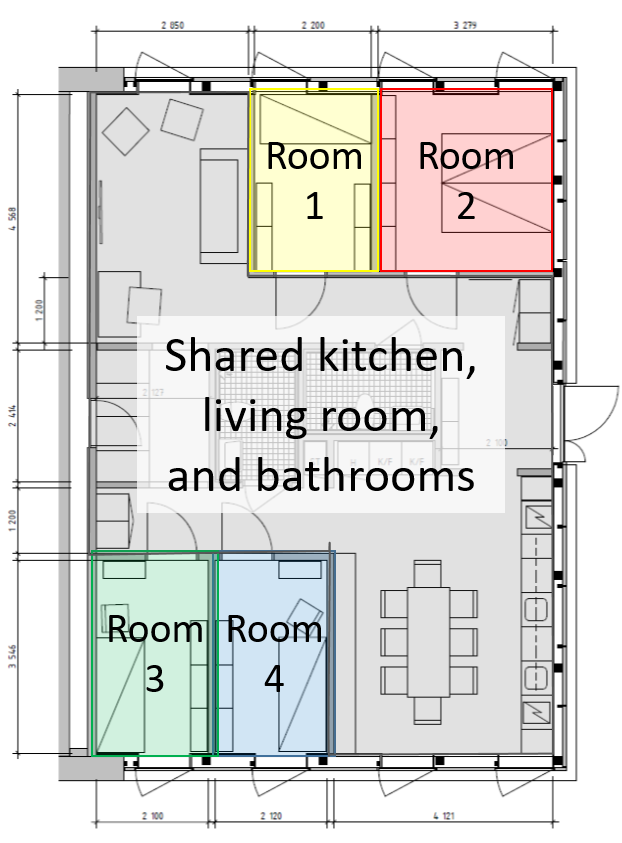}
&
\includegraphics[width=.25\textwidth]{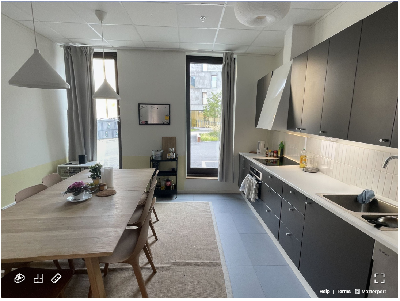}
\includegraphics[width=.25\textwidth]{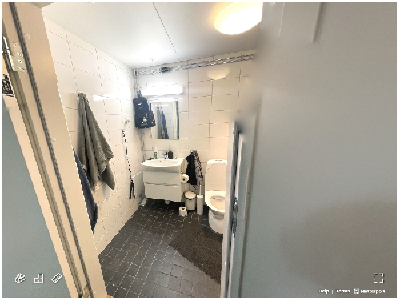}\newline
\includegraphics[width=.25\textwidth]{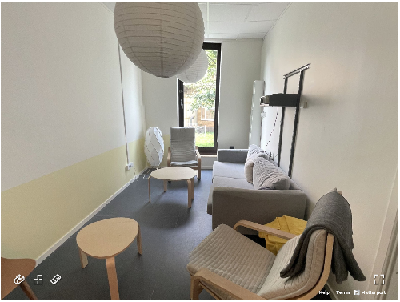}
\includegraphics[width=.25\textwidth]{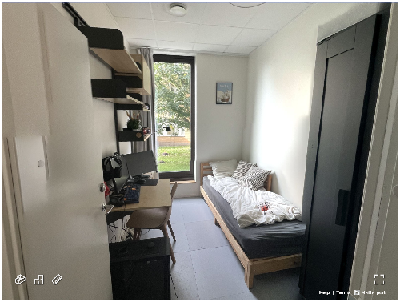}
\\
\small(a)  & \small(b) 
\end{tabular}
\caption{
(a): 
Illustration of KTH Live-In Lab used in this experimental campaign. The majority of the living area is covered by shared spaces, i.e., kitchen, living room and bathrooms; private spaces for the testbed occupants are limited to the bedrooms. 
(b): Indoor view of the KTH Live-In Lab.
}
\label{fig:TestbedEM}
\end{figure*}

The KTH Live-In Lab includes several building testbeds that range from students accommodations to lecture buildings; data collected from the sensors installed in the testbeds is stored and shared through the Live-In Lab datapool,
see \cite{Molinari2023Bottlenecks} for details. 
The building used in this preliminary campaign features an extended sensor network, advanced interaction capability with testbed occupants, and the possibility to redesign the testbed layout.
Figure~\ref{fig:TestbedEM} illustrates the KTH Live-In Lab layout, which comprises four individual rooms, two bathrooms, a kitchen, and a common area for an overall floor area of 300 $\text{m}^2$ (see Figure~\ref{fig:TestbedEM}(a), bottom right panel, and Figure~\ref{fig:TestbedEM}(b) for an indoor view of the rooms).  
Advanced sensing technologies 
are used to monitor indoor environment parameters such as indoor temperature, relative humidity, and CO$_2$, see \cite{Molinari2022LongTerm}. Additional sensors, like contact sensors, are deployed to detect, for instance, windows opening and occupancy to optimize the use of energy for heating and ventilation. Lighting sensors are used to study internal illuminance, maximize the use of daylight, and improve the light comfort.

KPIs (Key Performance Indicators) are considered as a key tool for measuring the performance of human behavior towards sustainability; however, monitoring a random set or large number of KPIs might be inefficient (see, e.g., \cite{mourtzis:2018}). It is important to select a small set of appropriate KPIs based on data availability, controllability, and its alignment with providing a clear view of all aspects of sustainable human behavior. 
According to \cite{Itard2008Building}, buildings and their occupants are responsible for a large part of energy consumption leading to a high impact on climate change, thus are important for sustainability. In general, buildings' energy consumption is evaluated by considering the energy required for heating, cooling, and domestic hot water, see \cite{ratajczak2021real}. Hence, in this research, we consider energy-related indicators to measure the level of sustainability of tenants of the KTH Live-In Lab:
\begin{itemize}
    \item Heating energy (kWh/m$^2$/year);
    \item Electricity use (kWh/dw);
	\item Domestic hot water (dm$^3$/day/person);
	\item Fresh water (dm$^3$/day/person).
\end{itemize}
The above mentioned KPIs are commonly used in the building sector. This facilitates their interpretability and the comparison with already existing references in literature. Note that some of the previously mentioned KPIs are affected by seasonality e.g., the heating energy is higher in winter than in summer. Moreover, they have high variability regarding reported average values in literature. For instance, electricity consumption varies with respect to the type of heating (e.g., electricity heating), the size of the dwelling, and the number of occupants in the apartments.

\subsection{The tenants}\label{sec:tenants}
\begin{figure*}[t]\centering
\includegraphics[width=1\textwidth]{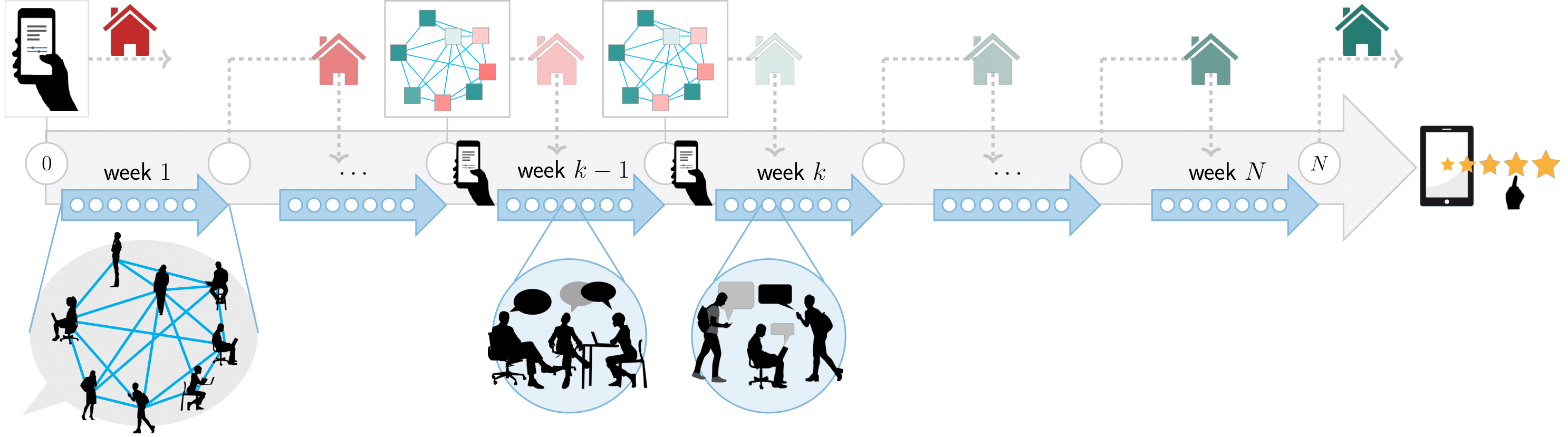}
\caption{Illustration of the longitudinal case study within the considered period of time, from week $0$ (start, reference state) to week $N$ (end of study), see Section~\ref{sec:timeplan} for a timeplan. A pre-study (week $0$) and a post-study (after week $N$) surveys are distributed to participants to obtain a reference state for behaviors and feedback on the experiment, respectively. The two arrow colors represent two time scales, days (light blue arrows) and weeks (gray arrow), capturing two different sources of social influence.
Social influence by \textit{communication}: During each week of the experimental campaign, tenants can (and are encouraged to) interact and exchange opinions with their neighbors in the social network, via in-person discussions, favored by the presence of common areas in the KTH Live-In Lab, and/or via messaging apps/group chats (bottom zoomed views).
Social influence by \textit{observation}: At the end of each week, tenants compile a questionnaire on past week behaviors regarding a set of sustainable practices (Table~\ref{tab:activities}, Section~\ref{sec:KPIs} for details), from which their sustainability scores are computed (top zoomed views). Subsequently, at the beginning of a new week, tenants can look at the average household sustainability score (defined in Section~\ref{sec:score}).}
\label{fig:longitudinal}
\end{figure*}

The tenants of the KTH Live-In Lab are mainly undergraduate students attending KTH Royal Institute of Technology, and residing at the KTH Live-In Lab full-time. 
The concept proposed in this paper is to represent the KTH Live-In Lab as a social network of interacting tenants (see Figure~\ref{fig:smart-home}(b)). In particular, a social network can be represented as a graph, whose nodes represent agents (or, equivalently, tenants), and an edge between a pair of nodes represents a social tie (or, cooperative/friendship relationship) among the corresponding agents.
In the experimental study, the tenants not only communicate and exchange opinions or beliefs among each other, but also interact with the KTH Live-In Lab (e.g., by receiving feedback on the household behavior, see Figure~\ref{fig:longitudinal}).

\section{Design of the case study}\label{sec:LiLSocial}
The implementation includes the design of a longitudinal experimental study that monitors sustainability practices (habits/behaviors that affect sustainability) of participants over a period of time, to detect any changes that might have occurred; Section~\ref{sec:timeplan} introduces the timeplan, Section~\ref{sec:KPIs} describes the sustainability indicators, and Section~\ref{sec:score} defines sustainability score for each tenant and average household sustainability score. Some final considerations on the design are discussed in Section~\ref{sec:discussion}.

\subsection{Timeplan}\label{sec:timeplan}
The study is divided into two main phases: a preparatory phase (winter 2022-2023), which we describe in details in the following sections and on which this paper focuses, and an execution phase (starting on spring 2023); the preparatory phase aims at testing the proposed approach before scaling it to a larger scale in the execution phase. The preparatory phase of this experimental study is conducted at the KTH Live-In Lab, described in Section~\ref{sec:KTH-LiL}, and the four occupants of the KTH Live-In Lab were asked to be involved for a time period spanning approximately 5 weeks. The aim is: (i) to test the general methodology and tools, and (ii) to assess the general approach through feedback from the participants.
Incentives for participation in the preparatory phase of the study include: refreshments during a Q\&A session organized before the beginning of the study and gift cards (associated with a lottery).

Data will be collected before, during, and after the experiment, see the illustration proposed in Figure~\ref{fig:longitudinal}. 

\textit{Before the experiment}. A pre-study survey is sent to participants to obtain demographic information, and understand their habits and how they perceive the environmental impact of their lifestyle. The aim is twofold: to avoid obtaining a biased sample, and to obtain a reference state (or, baseline data/initial conditions) that can later be used to analyze if and how there have been changes in behavior.\vspace{-0.1cm}

\textit{During the experiment}. The participants are asked to interact with each other and to compile weekly surveys ($\sim$ 10-15 min/week, see Section~\ref{sec:KPIs}), based on which their sustainability score is calculated (see Section~\ref{sec:score}). Afterwards, participants will receive feedback on the average household sustainability score (defined in Section~\ref{sec:score}). 
Moreover, measurements from sensors at apartment level are collected to understand the household consumption (see Section~\ref{sec:KTH-LiL}).
Again, the aim is twofold: to observe the dynamics of (interconnected) tenants’ sustainability score, and to investigate tenants' awareness of their consumption by comparing observed and self-reported data.\vspace{-0.1cm}

\textit{At the end of the experiment}. 
A meeting will be planned with the participants, whose purpose is to obtain feedback on the experimental study, including (but not limited to) time investment, ease of usage of the app utilized for the surveys, survey complexity, and perceived privacy concerns. These issues will be revised, and the execution phase of the study will be implemented thereafter.

\subsection{Sustainability indicators}\label{sec:KPIs}
In this section we discuss the KPIs we have introduced in our study in order to provide a measure of the sustainability of human behaviors in the building context.
We define \emph{environmentally significant behavior} as behavior that «changes the availability of materials or energy from the environment or alters the structure and dynamics of ecosystems or the biosphere itself» (\emph{impact-oriented} definition), or «is undertaken with the intention to change (normally, to benefit) the environment» (\emph{intent-oriented} definition) \cite[p. 408]{Stern2000Toward}).
To measure tenant's behaviors we resort to self-reports in the form of weekly surveys, and we design our questionnaire around 17 sustainability practices (or activities) grouped in the following 5 lifestyle dimensions:
\begin{enumerate}
    \item \emph{Conservation or Resources}: consumption of electricity, domestic hot water, fresh water, heating energy;
    \item \emph{Consumption}: waste production and recycling, electronic waste production, clothing choices;
    \item \emph{Food}: consumption of beef, pork, poultry, dairy;
    \item \emph{Transportation or Mobility}: taking public transportation, and walking or cycling instead of driving;
    \item \emph{Environmental citizenship}: membership in environmental/conservation organizations and frequency of talking to others about their environmental behavior.
\end{enumerate}
Table~\ref{tab:activities} presents the complete list of activities, where the number 17 is the result of a trade-off process in an effort to keep the survey simple and not too time-consuming, but still informative and spanning all dimensions. Each activity is formulated as a choice between two complementary/alternative options.

We distribute two types of surveys, at the end and beginning of each week.
In the surveys distributed at the end of each week, tenants are asked to report their past \textit{actions} (``In the past week, how often did you\dots'') and \textit{beliefs} in terms of environmental responsibility vs. burden caused by each activity.
In the surveys distributed at the beginning of each week, tenants are asked to report their \textit{intention} in terms of sustainability score for the coming week, after observing the average household sustainability score.

The surveys are formulated using Qualtrics software\footnote{\copyright 2022 Qualtrics, Provo, UT, USA. \url{https://www.qualtrics.com}}, with anonymized links made available to the participants. Behaviors are measured using a Visual Analogue Scale (VAS) where, for each statement, participants are asked to select a value along a continuous line between two end points, from $0$ to $100$, representing opposite values (e.g., $0=$ strongly disagree, $50=$ neutral, $100=$ strongly agree). 

Let $n$ be the total number of tenants in the household, and $N$ be the total number of weeks the experiment will run. Then, the action of tenant $i=1,\dots,n$ regarding activity $q=1,\dots,17$ of Table~\ref{tab:activities} at (the end of) week $k=0,1,\dots,N$ is denoted by
\begin{equation}
    y_{i,q}(k) \in [0,100]. 
    \label{eqn:action-tenant}
\end{equation}
In eq.~\eqref{eqn:action-tenant}, $k=0$ represents the start of the campaign and $y_{i,q}(0)$ the reference state obtained from the pre-study survey, see Section~\ref{sec:timeplan}.

\subsection{Sustainability score}\label{sec:score}
For the definition of sustainability scores we use the data collected in the survey on actions of tenants, see eq.~\eqref{eqn:action-tenant}.
\begin{defn}
The \emph{sustainability score} of each tenant $i=1,\dots,n$, where $n$ is the total number of tenants in the household, at week $k=0,1,\dots,N$ is given by
\begin{equation}
    \score_i(k) = \sum_{q=1}^{17} y_{i,q}(k) \in [0,1700],
    \label{eqn:score-tenant}
\end{equation}
where $y_{i,q}(k)\in [0,100]$ represents the actions-behavior reported by tenant $i$ in week $k$ w.r.t. the $q$-th activity, see eq.~\eqref{eqn:action-tenant}.
The \emph{average household sustainability score} at week $k=0,1,\dots,N$ is given by
\begin{equation}
    \hscore(k) = \frac{1}{n} \sum_{i=1}^n \score_i(k) \in [0,1700].
\label{eqn:score-house}
\end{equation}
\end{defn}

\subsection{Discussion}\label{sec:discussion}
Two limitations regarding the KPIs should be mentioned. 
The energy-related KPIs for the KTH Live-In Lab (Section~\ref{sec:KTH-LiL}) evaluate an \textit{overall} measure of occupants' desire towards sustainability. However, in reality, occupants may have a mixed level of incentive for different activities and behaviors which cannot be captured by the selected KPIs. For instance, people may be more motivated to turn off the lights to reduce electricity consumption than to keep the windows closed during the winter time to avoid excessive energy consumption. Future studies will take detecting an activity change into account, starting from observing window opening behavior of occupants during the winter through existing sensors in KTH Live-In Lab (see, e.g., \cite{hong2016advances}). 
Another limitation is the use of self-reports (surveys) as indicators of behavior performance (Section~\ref{sec:KPIs}), which proves restrictive as it needs to rely on participants' ``honesty'' and awareness, instead of actual observations, to determine behavior. However, this also gives us the opportunity to evaluate the awareness of tenants regarding the behavior related to resources (Table~\ref{tab:activities}), by comparing the surveys' data (Section~\ref{sec:KPIs}) with the data collected at KTH Live-In Lab (Section~\ref{sec:KTH-LiL}).

Finally, besides exploring the role of social influence, the ambitious aim behind gathering data on \textit{beliefs} and \textit{intention} (Section~\ref{sec:KPIs}) is to explore causes/motivations of potential discrepancies between actions and beliefs, i.e., the so-called value-action gap (\cite{Kollmuss2002Mind}), repeatedly observed in the sustainability context.
The data collected is to be used to support the analysis and evaluation of theoretical models for decision-making over social networks, and ultimately to improve the building energy gap (i.e., discrepancy between expected and measured energy consumption, \cite{Molinari2022LongTerm}).

\section{Preliminary results}\label{sec:results}
\begin{figure*}\centering
\setlength{\tabcolsep}{0pt}
\begin{tabular}[t]{G{.66\textwidth}G{.34\textwidth}}
\includegraphics[height=4.4cm]{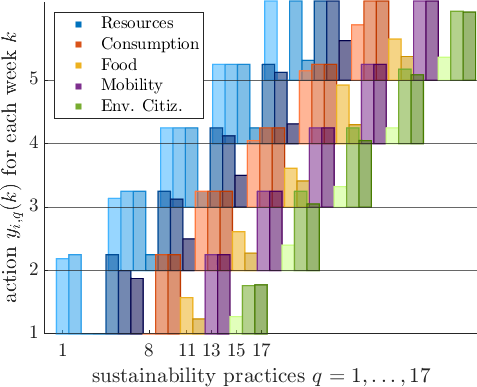}\;  \includegraphics[height=4.7cm]{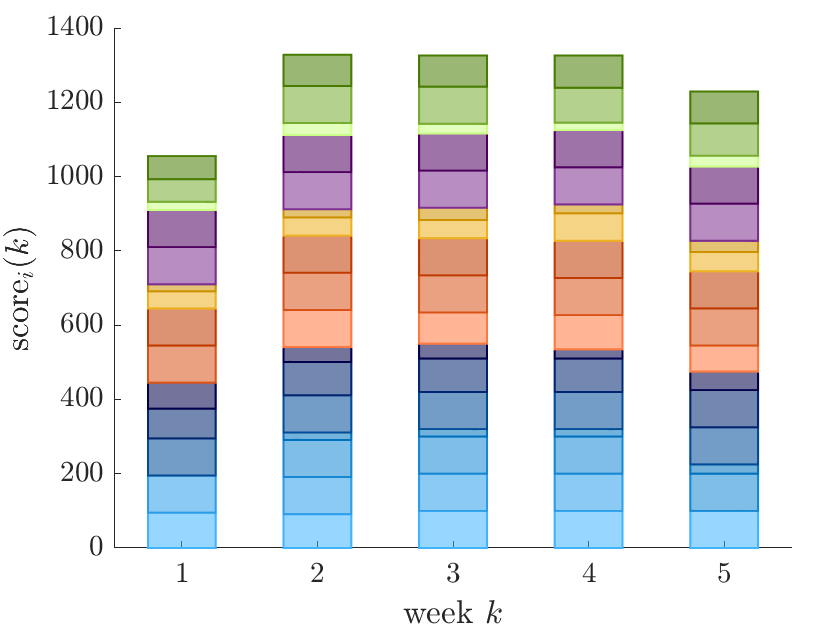}&
\includegraphics[height=4.65cm]{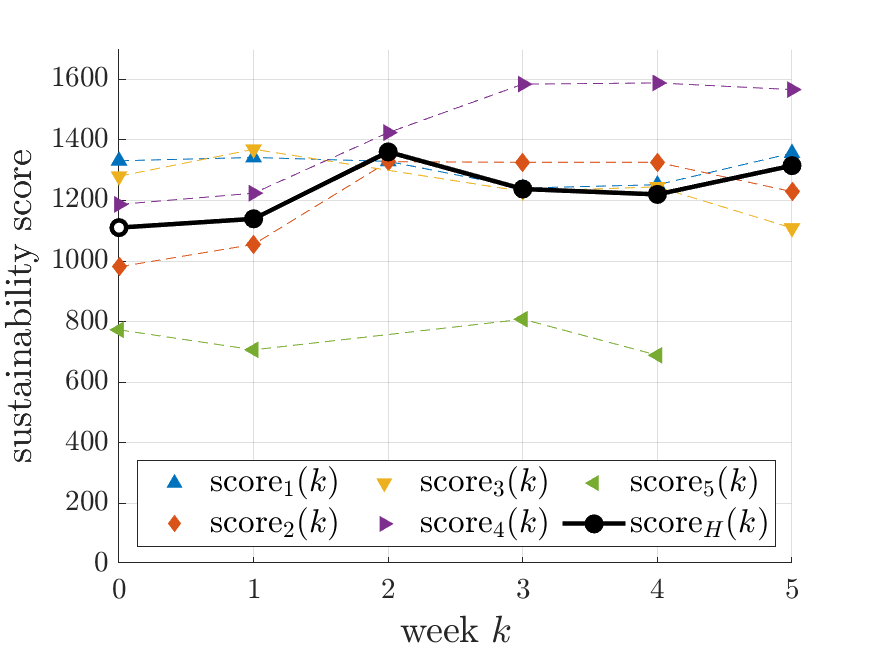}
\\
(a) & (b)
\end{tabular}
\caption{(a): Actions ($y_{i,q}$, eq.~\eqref{eqn:action-tenant}) related to sustainability practices (Table~\ref{tab:activities}) (left panel) and sustainability score ($\score_i$, eq.~\eqref{eqn:score-tenant}) (right panel) of one of the tenants of the KTH Live-In Lab in a 5-week period. (b): Sustainability scores of all tenants and average household sustainability score ($\score_H$, eq.~\eqref{eqn:score-house}) in a 5-week period.}
\label{fig:results}
\end{figure*}

The experimental study run in the period between beginning of February and end of March 2023 and, while a meeting with the tenants is currently being planned, preliminary feedback indicates positive responses towards the study and tools used. Fig.~\ref{fig:results} reports preliminary data, collected solely through the surveys related to actions of tenants w.r.t. the 17 sustainability practices (color-coded in Fig.~\ref{fig:results}(a) according to the 5 dimensions in Table~\ref{tab:activities}) over a 5-week period. 
Even if the small sample size and short monitoring period limit the extent of evidence regarding the effect of social influence, 
Fig.~\ref{fig:results}(b) shows an increase (between week $0$ and week $5$) in the average household sustainability score, $\hscore(k)$. Future works will be dedicated to a complete data analysis, including data collected at KTH Live-In Lab (Section~\ref{sec:KTH-LiL}) and data on tenants' beliefs (Section~\ref{sec:KPIs}).

\section{Conclusions}\label{sec:conclusion}
In this paper the preparatory phase of an experimental study on the effect of social influence in behavioral changes towards sustainability is presented, as means to improve our methodology before running the experiment at a larger scale. The concept the experimental design is based on consists of representing a smart home as a social network, and of investigating how different sources of social influence affect the environmentally significant behavior of the interconnected tenants.
A critical aspect to move from preparatory to execution phase is to evaluate the feeling of social belonging of the tenants in the context of sustainable practices. In other words, determine if the approach proposed, which promotes discussions among neighbors and provides feedback on sustainability scores, is enough to establish a sense of belonging and connectedness.
To boost participants/students engagement, we phrase the experiment (incl., survey practice and sustainability score) as a learning, collaborative, and inclusive opportunity for both parties, research team and participants.


\begin{thebibliography}{40}
\providecommand{\natexlab}[1]{#1}
\providecommand{\url}[1]{\texttt{#1}}
\providecommand{\urlprefix}{URL }
\expandafter\ifx\csname urlstyle\endcsname\relax
  \providecommand{\doi}[1]{doi:\discretionary{}{}{}#1}\else
  \providecommand{\doi}{doi:\discretionary{}{}{}\begingroup
  \urlstyle{rm}\Url}\fi

\bibitem[{Anderson and Ye(2019)}]{Anderson2019Recent}
Anderson, B.D.O. and Ye, M. (2019).
\newblock {Recent advances in the modelling and analysis of opinion dynamics on influence networks}.
\newblock \emph{Int J Autom. Comput.}, 16,
  129--149. 
  

\bibitem[{Annaswamy et~al.(2022)Annaswamy, Khargonekar, Lamnabhi-Lagarrigue,
  and Spurgeon}]{CPHS2022}
Annaswamy, A.M. et~al. (2022).
\newblock \emph{{Cyber-physical-human systems: fundamentals and applications}}.
\newblock Wiley, in press.

\bibitem[{Bernardo et~al.(2021)Bernardo, Wang, Vasca, Hong, Shi, and
  Altafini}]{Bernardo2021Paris}
Bernardo, C. et~al. (2021).
\newblock {Achieving consensus in multilateral international negotiations: The
  case study of the 2015 Paris Agreement on climate change}.
\newblock \emph{Science Advances}, 7(51), 1--17.

\bibitem[{Bibri and Krogstie(2017)}]{Bibri2017Smart}
Bibri, S.E. and Krogstie, J. (2017).
\newblock {Smart sustainable cities of the future: An extensive
  interdisciplinary literature review}.
\newblock \emph{Sustainable Cities and Society}, 31, 183--212.

\bibitem[{Das et~al.(2020)Das, Konstantakopoulos, Manasawala, Veeravalli, Liu,
  and Spanos}]{das-2020do}
Das, H.P. et~al. (2020).
\newblock Do occupants in a building exhibit patterns in energy consumption?
  analyzing clusters in energy social games.
\newblock In \emph{NeurIPS 2020 Workshop on Tackling Climate Change with
  Machine Learning}.

\bibitem[{Fontan and Altafini(2021)}]{Fontan2021Signed}
Fontan, A. and Altafini, C. (2021).
\newblock {A signed network perspective on the government formation process in
  parliamentary democracies}.
\newblock \emph{Scientific Reports}, 11(5134).

\bibitem[{Frederiks et~al.(2015)Frederiks, Stenner, and
  Hobman}]{Frederiks2015Household}
Frederiks, E.R., Stenner, K. and Hobman, E.V. (2015).
\newblock {Household energy use: applying behavioral economics to understand
  consumer decision-making and behavior}.
\newblock \emph{Renewable Sustainable Energy Rev.}, 41, 1385--1394.

\bibitem[{{Government Offices of Sweden}(2016)}]{web:Sweden-sustainable}
{Government Offices of Sweden}.
\newblock (2016). 
  \href{https://www.government.se/government-policy/the-global-goals-and-the-2030-Agenda-for-sustainable-development/goal-11-sustainable-cities-and-communities}{government.se/ government-policy/the-global-goals-and-the-2030-Agenda-for-sustainable-development/goal-11-sustainable-cities-and-communities}.


\bibitem[{Hamilton and Rapf(2020)}]{Hamilton2020GlobalStatus}
Hamilton, I. and Rapf, O. (2020).
\newblock {Executive summary of the 2020 global status report for buildings and
  construction}.
\newblock \emph{Global Alliance for Buildings and Construction}.

\bibitem[{Hong et~al.(2016)Hong, Taylor-Lange, D’Oca, Yan, and
  Corgnati}]{hong2016advances}
Hong, T. et~al. (2016).
\newblock Advances in research and applications of energy-related occupant
  behavior in buildings.
\newblock \emph{Energy and buildings}, 116, 694--702.

\bibitem[{IEA(2019)}]{IEA2019}
IEA.
\newblock (2019). Total energy supply by source, Sweden.
  \href{https://www.iea.org/countries/sweden}{iea.org/countries/sweden}.

\bibitem[{IEA(2022)}]{IEA2022}
IEA.
\newblock (2022). Buildings. 
  \href{https://www.iea.org/reports/buildings}{iea.org/reports/buildings}.

\bibitem[{Intille(2006)}]{Intille2006SmartPeople}
Intille, S. S. (2006).
\newblock {The goal: smart people, not smart homes}.
\newblock In \emph{Int. Conf. Smart Homes Health Telem.}, 3--6.

\bibitem[{Intille et~al.(2006)}]{Intille2006LiL}
Intille, S. S. et~al. (2006).
\newblock{
Using a Live-In laboratory for ubiquitous computing research}.
\emph{Pervasive Computing}, 349--365.

\bibitem[{Itard et~al.(2008)Itard, Meijer, Vrins, and
  Hoiting}]{Itard2008Building}
Itard, L. et~al. (2008).
\newblock {Building Renovation and Mod-ernisation in Europe: State of the art
  review}.
\newblock Technical report, OTB Research Institute for Housing, Urban and
  Mobility Studies, Delft University of Technology, Delft.

\bibitem[{Karvonen et~al.(2021)Karvonen, Cvetkovic, Herman, Johansson,
  Kjellstr{\"{o}}m, Molinari, and Skoglund}]{Karvonen2021Urban}
Karvonen, A. et~al. (2021).
\newblock {The ‘New Urban Science’: towards the interdisciplinary and
  transdisciplinary pursuit of sustainable transformations}.
\newblock \emph{Urban Transform.}, 3(1).

\bibitem[{Knutsson and Marx(2016)}]{Knutsson2016Habitation}
Knutsson, J. and Marx, C. (2016).
\newblock {the Habitation Lab As a Platform for Sustainable Innovations in the
  Built Environment: a Case Study}.
\newblock In \emph{Proceedings Int. Multidiscip. Sci. GeoConference SGEM}, 2, 317--324.


\bibitem[{Kollmuss and Agyeman(2002)}]{Kollmuss2002Mind}
Kollmuss, A. and Agyeman, J. (2002).
\newblock {Mind the Gap: Why do people act environmentally and what are the
  barriers to pro-environmental behavior?}
\newblock \emph{Environmental Education Research}, 8(3), 239--260.

\bibitem[{Lamnabhi-Lagarrigue et~al.(2017)Lamnabhi-Lagarrigue, Annaswamy,
  Engell, Isaksson, Khargonekar, Murray, Nijmeijer, Samad, Tilbury, and Van~den
  Hof}]{Lamnabhi-Lagarrigue2017Systems}
Lamnabhi-Lagarrigue, F. et~al. (2017).
\newblock {Systems {\&} Control for the future of humanity, research agenda:
  Current and future roles, impact and grand challenges}.
\newblock \emph{Annual Reviews in Control}, 43, 1--64.

\bibitem[{Markle(2013)}]{Markle2013PEBS}
Markle, G.L. (2013).
\newblock {Pro-Environmental Behavior: Does It Matter How It's Measured?
  Development and Validation of the Pro-Environmental Behavior Scale (PEBS)}.
\newblock \emph{Human Ecology}, 41(6), 905--914.


\bibitem[{Molinari et~al.(2023)}]{Molinari2023Bottlenecks}
Molinari, M., Vogel, J.A., Rolando, D., Lundqvist P.L. (2023).
\newblock {Using living labs to tackle innovation bottlenecks: the KTH Live-In Lab case study}.
\newblock \emph{Applied Energy}, 338, 120877--120877.

\bibitem[{Mourtzis et~al.(2018)Mourtzis, Papatheodorou, and
  Fotia}]{mourtzis:2018}
Mourtzis, D., Papatheodorou, A. and Fotia, S. (2018).
\newblock Development of a key performance indicator assessment methodology and
  software tool for product-service system evaluation and decision-making
  support.
\newblock \emph{J. Computing and Information Science in Engineering}, 18(4).

\bibitem[{Peng et~al.(2012)Peng, Yan, Wu, Wang, Zhou, and
  Jiang}]{Peng2012Residential}
Peng, C. et~al. (2012).
\newblock {Quantitative description and simulation of human behavior in
  residential buildings}.
\newblock \emph{Building Simulation}, 5(2), 85--94.

\bibitem[{Prati et~al.(2017)Prati, Albanesi, and
  Pietrantoni}]{Prati2017Longitudinal}
Prati, G., Albanesi, C. and Pietrantoni, L. (2017).
\newblock {The interplay among environmental attitudes, pro-environmental
  behavior, social identity, and pro-environmental institutional climate. A
  longitudinal study}.
\newblock \emph{Environ. Education Research}, 23(2), 176--191.

\bibitem[{Proskurnikov and Tempo(2017)}]{Proskurnikov2017Tutorial}
Proskurnikov, A.V. and Tempo, R. (2017).
\newblock {A tutorial on modeling and analysis of dynamic social networks. Part
  I}.
\newblock \emph{Annual Reviews in Control}, 43, 65--79.

\bibitem[{Ratajczak et~al.(2021)Ratajczak, Michalak, Narojczyk, and
  Amanowicz}]{ratajczak2021real}
Ratajczak, K. et~al. (2021).
\newblock Real domestic hot water consumption in residential buildings and its
  impact on buildings’ energy performance—case study in Poland.
\newblock \emph{Energies}, 14(16), 5010.

\bibitem[{Ritchie(2020)}]{Ritchie2020}
Ritchie, H.
\newblock (2020). Sector by sector: where do global greenhouse gas emissions
  come from? \href{https://ourworldindata.org/ghg-emissions-by-sector}{ourworldindata.org/ghg-emissions-by-sector}.

\bibitem[{Rolando et~al.(2022)Rolando, Mazzotti~Pallard, and
  Molinari}]{Molinari2022LongTerm}
Rolando, D., Mazzotti~Pallard, W. and Molinari, M. (2022).
\newblock {Long-term evaluation of comfort, indoor air quality and energy
  performance in buildings: The case of the KTH Live-In Lab Testbeds}.
\newblock \emph{Energies}, 15(14), 4955.

\bibitem[{Sasic~Kalagasidis et~al.(2018)Sasic~Kalagasidis, Hagy, and
  Marx}]{Kalagasidis2017HSB}
Sasic~Kalagasidis, A., Hagy, S. and Marx, C. (2018).
\newblock {The HSB Living Lab harmonization cube}.
\newblock \emph{Informes de la Construcci{\'{o}}n}, 69(548), 224--12.

\bibitem[{Steg and Vlek(2009)}]{Steg2009Encouraging}
Steg, L. and Vlek, C. (2009).
\newblock {Encouraging pro-environmental behaviour: An integrative review and
  research agenda}.
\newblock \emph{J. Environ. Psychology}, 29(3), 309--317.

\bibitem[{Stern(2000)}]{Stern2000Toward}
Stern, P.C. (2000).
\newblock {New Environmental Theories: Toward a Coherent Theory of
  Environmentally Significant Behavior}.
\newblock \emph{Journal of Social Issues}, 56(3), 407--424.

\bibitem[{Sultan et~al.(2020)Sultan, Spanos, Goh, Ping, and Sapar}]{Spanos2020}
Sultan, Z. et~al. (2020).
\newblock Living Lab Integrates IEQ Technologies in Singapore.
\newblock \emph{ASHRAE J.}, 68–71.

\bibitem[{{Swedish Energy Agency}(2022)}]{web:Swedish-energy}
{Swedish Energy Agency}.
\newblock (2022).
  \href{https://www.energimyndigheten.se/en/sustainability/every-kilowatt-hour-kwh-counts/}{energimyndigheten.se/ en/sustainability/every-kilowatt-hour-kwh-counts}.

\bibitem[{Tian and Wang(2018)}]{Tian2018Opinion}
Tian, Y. and Wang, L. (2018).
\newblock {Opinion dynamics in social networks with stubborn agents: An
  issue-based perspective}.
\newblock \emph{Automatica}, 96, 213--223.


\bibitem[{Wang et~al.(2021)Wang, Bernardo, Hong, Vasca, Shi, and
  Altafini}]{Wang2021Concatenated}
Wang, L. et~al. (2021).
\newblock {Achieving consensus in spite of stubbornness: time-varying
  concatenated Friedkin-Johnsen models}.
\newblock In \emph{60th IEEE Conference on Decision and Control}, 
  4964--4969.

\bibitem[{Wilson and Dowlatabadi(2007)}]{Wilson2007Residential}
Wilson, C. and Dowlatabadi, H. (2007).
\newblock {Models of decision making and residential energy use}.
\newblock \emph{Annual Review of Environment and Resources}, 32, 169--203.


\bibitem[{Zino et~al.(2020)Zino, Ye, and Cao}]{Zino2020Twolayer}
Zino, L., Ye, M. and Cao, M. (2020).
\newblock {A two-layer model for coevolving opinion dynamics and collective
  decision-making in complex social systems}.
\newblock \emph{Chaos}, 30(8).

\end{thebibliography}

\end{document}